\documentclass[12pt]{iopart}

\usepackage{iopams}
\usepackage{graphicx}
\begin{document}

\title{Identification of clusters of investors from their real trading activity in a financial market }

\author{Michele Tumminello$^{1,2}$, Fabrizio Lillo$^{1,3,4}$, Jyrki Piilo$^5$, Rosario N. Mantegna$^1$}
\address{$^1$ Dipartimento di Fisica, Universit\`a di Palermo, Viale delle Scienze Ed. 18, I-90128 Palermo, Italy}
\address{$^2$ Department of Social and Decision Sciences, Carnegie Mellon University, Pittsburgh, PA15213, USA}
\address{$^3$ Scuola Normale Superiore di Pisa, Piazza dei Cavalieri 7, 56126 Pisa, Italy}
\address{$^4$ Santa Fe Institute, 1399 Hyde Park Road, Santa Fe NM 87501, USA}
\address{$^5$ Turku Centre for Quantum Physics, Department of Physics and Astronomy, University of Turku, 
FI-20014 Turun yliopisto, Finland}

\ead{rn.mantegna@gmail.com}

\begin{abstract}
 We use statistically validated networks, a recently introduced method to validate links in a bipartite system, to identify clusters of investors trading in a financial market. Specifically, we investigate a special database allowing to track the trading activity of individual investors of the stock Nokia. We find that many statistically detected clusters of  investors show a very high degree of synchronization in the time when they decide to trade and in the trading action taken. We investigate the composition of these clusters and we find that several of them show an over-expression of specific categories of investors.    
\end{abstract}

\maketitle
\tableofcontents

\newpage

\section{Introduction}

 Financial markets are complex systems where the simultaneous activity of a huge number of investors performs the task of finding the correct price of an asset through the action of trading. The way in which this collective task is obtained is only partially understood. Theoretical and computational models of investors trading in a financial markets are very helpful in reproducing stylized facts and allow to investigate how specific cognitive assumptions and investment strategies affect the price dynamics \cite{Hommes2006,Samanidou2007}. Theoretical and computational models often classify investors in stylized classes such as the ones of fundamentalists and chartists \cite{Frankel1990,Chiarella1992,Kirman1993,Lux1995,Brock1998,Lux1999}  sometime specialized in contrarian \cite{Chan1988} and momentum \cite{Chan1996} investors. Theoretical and computational models also distinguish between informed and uninformed trading action, for example see \cite{Grossman1976,Roll1984,Kyle1985,Hasbrouck2007}, up to the limit case of considering the presence of only noise driven zero intelligence traders \cite{Gode1993,Farmer2005}.
In the above cited studies, the assumptions on the presence of different classes of investors are motivated by theoretical considerations, results of surveys and direct investigations of the trading profile of classes of investors  (see \cite{Nofsinger1999,Choe1999,Griffin2003} for some examples of these investigations).

Even if theoretical and computational models take into account some type of agent's heterogeneity, analytical tractability or the need to keep low the number of model parameters force researchers to introduce a small number of groups of investors characterized by a specific type of strategy. Without an empirical verification of the underlying assumptions, the assessment of the real amount of heterogeneity present in a market and the detection of its role in the price formation dynamics lack an empirical support. There are at least two reasons why such empirical investigations are difficult to be realized. First, due to confidentiality reasons, it is very difficult to have access to data allowing to track the trading activity of a large set of individual investors for a long period of time. Second, even when the data are available, the identification of groups of investors trading in similar way is a complicated data mining task. 

In this paper we make a first step in this direction by employing a recently developed and powerful data mining technique, termed Statistically Validated Networks (SVN) \cite{Tumminello2011}, for the analysis of a very special database, namely a database allowing to track the trading activity of individual investors of Finnish stocks.  With our approach we are able to identify groups of investors that trade in a very similar way over extended periods of time. This commonality of behavior can be due to the use of very similar trading strategies and can be seen as a strong form of herding. One of the most surprising results is that in some groups we find  a very high degree of synchronization in agent's trading activity, both in terms of when they decide to trade (as opposed to maintain their position) and in terms of the specific activity (i.e. buy, sell, or buying and selling approximately the same 
amount of shares in a given day) performed in a given day. In this paper we will not investigate why the identified groups follow a specific trading patterns, i.e. we will not attempt here a reverse engineering approach to infer strategies from trading activity of investors. This is the topic of a forthcoming paper \cite{Tumminello2012}. Our main focus here is in the identification and compositional characterization of the identified clusters of investors.  

The task of identifying groups of investors and infer their strategies and interactions from empirical data is receiving an increasing interest in recent years. 
Some papers \cite{Lillo2008,Vaglica2008,Moro2009,Toth2010, Carollo2011,Toth2011} have investigated databases where it is possible to track the trading behavior of market members of the exchange. Members are credit entities and investment firms which  are the only firms entitled to trade directly. Therefore they trade on behalf of a large number of investors. Despite this fact, recent studies have shown that, probably due to a customer specialization, market member data allows to identify trading strategies, such as order splitting \cite{Vaglica2008,Moro2009,Toth2010}, liquidity provision \cite{Toth2011}, and contrarian or momentum trading \cite{Lillo2008}. In particular in this last study authors have performed an analysis of the linear correlation matrix of the trading activity of market members of the Spanish Stock Exchange in order to identify groups of investors (market members in this case). It is important to stress that, as will be clear in the following, this approach cannot be pursued with the Finnish data of individual investors investigated here. The main reason is the extreme heterogeneity in the trading activity of individual investors (heterogeneity is not so significant for market member data). For this reason in this paper we use the more sophisticated SVN to identify clusters of investors.

Other studies have had access to databases with the resolution of the individual investors (see for example \cite{Barber2008} for the profit analysis of Taiwanese investors, \cite{Kirilenko2010} for the analysis of the Flash Crash of May 6, 2010 or \cite{Feiren2011} for the investigation of order splitting for individual investors).  The database used in this paper  has been investigated extensively by Grinblatt and Keloharju in a series of studies \cite{Grinblatt2000,Grinblatt2009} on the trading profile of individual and institutional investors, and on behavioral aspects of individual investors. However to the best of our knowledge, this is the first study that attempts an unsupervised identification of groups of individual investors in a financial market.
 
 The paper is organized as follows. In Section II we describe the system and the special database investigated in this study. In Section III we describe the categorical variables used to characterize the trading activity of the investors and we introduce the bipartite system under investigation. Section IV describes the statistically validated networks of investors and Section V investigates the clusters detected in the statistical validated networks. Finally, Section VI concludes. 

\section{Data}\label{data}

We investigate the trading activity of institutional and individual single investors by using a special database maintained by the Euroclear Finland (previously Nordic Central Securities Depository Finland). 
The database is the central register of shareholdings for Finnish stocks and financial assets in the Finnish Central Securities depository (FCSD). Practically all major publicly traded Finnish companies have joined the register. The register reports the shareholdings in FCSD stocks of all Finnish investors and of all foreign investors asking to exercise their vote right, both retail and institutional. The database records official ownership of companies and financial assets and the trading records are updated on a daily basis according to the Finnish Book Entry System. The records include all the transactions, executed in worldwide stock exchanges and in other venues, which change the ownership of the assets.

The database classifies investors into six main categories: non-financial 
corporations, financial and insurance corporations, general governmental organizations, non-profit institutions, households, and foreign organizations. 
The database is collected since January 1st, 1995.  We have access to the database for the period 1995-2008, under a special agreement with Euroclear Finland. In this paper we investigate the trading activity of the Nokia stock, which is the most capitalized stock in the Finnish stock market. Note that the database covers transactions of Nokia in all financial markets where this company is listed. However,  
while the database contains very detailed information about the Finnish domestic investors, foreign investors can choose to use nominee registration. In this case, the investor's book entry account provider aggregates all the transactions from all of its accounts, and a single nominee register coded identity contains the holdings of several foreign investors. 
\footnote{If an institution can trade both for itself and also on behalf of nominee registered investors, we split its trading activity in two distinct IDs, one regarding its activity as a Finnish investor and one when it trades for nominee registered investors.}
This means that our results describe in detailed way the actions of all the Finnish domestic investors and those foreign investors who do not use nominee registration, while a very small fraction of the coded identities correspond to aggregated ownership.

We consider the set of investors trading the Nokia stock during the period of time from 19 October 1998  to 29 December 2003 (a set of 1,300 daily records) and we investigate all the market transactions performed by them. The total number of investors is $164,130$ and the total number of transactions is $18,313,376$. The left part of Table \ref{TableDescr1} gives the number of investors, the number of transactions, and the traded volume for the six categories. It also gives these numbers separating nominee registered and non nominee registered investors. 

\begin{table}
\caption{\label{TableDescr1} Summary of the number of investors (\# ids), the number of transactions ($N$), and the exchanged volume ($V$, in millions of shares) for the entire set (left column) of Nokia investors in the period Oct. 10, 1998 - Dec. 29, 2003 and for the restricted set (right column) of active investors investigated in the paper (see text for a detailed description of the set construction). The investors are divided in the six categories (top part) or between nominee and non nominee registered. Note that the total volume in the table is twice the traded volume, because each transactions is counted both for the buyer and for the seller.}
\begin{center}
\begin{tabular}{l|rrr|rrr}
  \hline
  ~	&	~	& Entire set & ~ & ~ & Restricted set & ~ \\
  \hline
  Category				& \# ids&$N$~~~ 	&$V$~~ 	&\# ids	&$N$~~~	&$V$~~	\\
  \hline
 Non financial corporations 	&9,298	&	 1,516,587& 8,996		&1,570	&1,464,776	&8,932	\\
 Financial and insurance 		&434	&	14,761,690&179,281	&206	&14,759,603	&179,272	\\ 
 Governamental			&144	&	        21,313& 618		&75		&20,462		&615	\\
 Non profit					&1,119	&	       25,955&276		&99		&17,683		&262	\\
 Households				&151,493	&	  1,646,482&1,083		&9,326	&1,002,918	&916	\\
 Foreign organizations		&1,642	&	     341,349&2,754		&109	&333,547		&2,680	\\
 \hline
Total						&164,130	&18,313,376	&193,008		&11,385	&17,598,989	&192,677	\\
\hline
\hline
Nominee registered			& 52			&	13,118,319 &	174,753 & 40 		& 13,118,203 & 174,738 \\
Non nominee registered		& 164,078		&	  5,195,057 &      18,255 & 11,345	& 4,480,786   & 17,939 \\	
\hline
\end{tabular}
\end{center}
\end{table}

\begin{figure}[t]
\begin{center}
\includegraphics[scale=0.5]{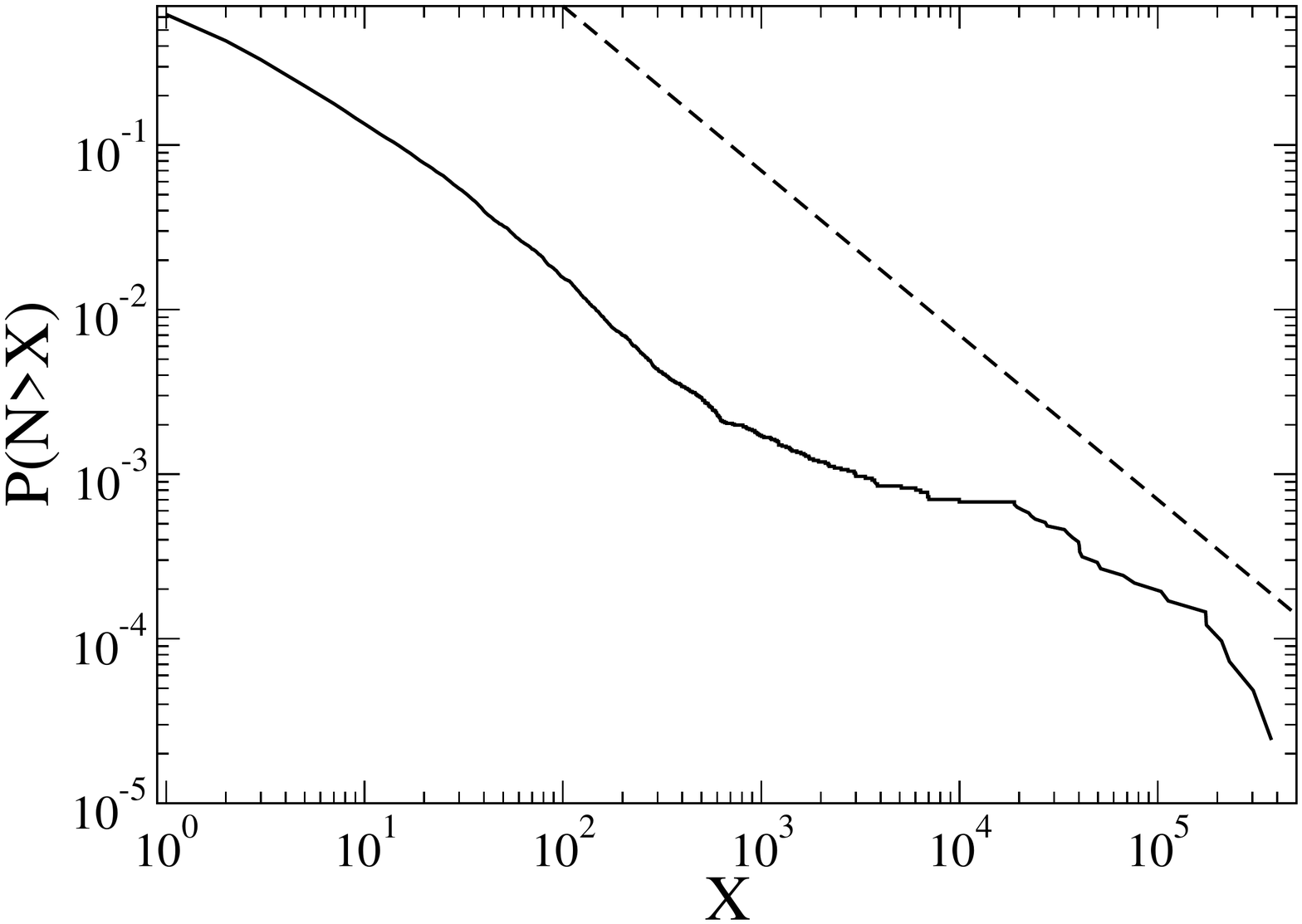}
\caption{Cumulative probability density, $P(N>X)$, of the number of market transactions $N$ performed by the $41,250$
investors trading the Nokia stock 
during 2003. The dashed line is a power law with slope $-1$ and serves as a guide for the eye. }
\label{fig1}
\end{center}
\end{figure}

Investment decisions of single investors are characterized by a huge heterogeneity with respect to: i) individual, collective or institutional nature, ii) investment size, iii) investment time horizon, iv) class of trading strategies, and v) information sources and processing capabilities. Therefore, the investigation of institutional and individual investors immediately faces the limitations imposed by the investors' heterogeneity.  As an example of the degree of heterogeneity in trading activity observed in a financial market, we show in Fig.~\ref{fig1} the cumulative probability density function of the number of market transactions for investors
trading the Nokia stock during 2003. The cumulative distribution shows a high degree of heterogeneity. The number of transactions performed by investors ranges from $1$ to $1,549,871$. In the region of low number of transactions ($N\lesssim1000$) 
the cumulative probability density is roughly approximated by a power-law with an exponent close to $-1$ as in the famous Zipf's law \cite{Zipf} observed for other size distributions such as cities population or firms size \cite{Simon,Axtell}. However, it is worth noting that, differently than in the classic cases of observation of a Zipf's law, the best agreement with a power-law behavior with an exponent close to $-1$ is not observed in the tail of the cumulative distribution and therefore for most active investors but rather for investors characterized by a number of transactions ranging from one to roughly one thousand. The figure also shows that many investors made very few transactions, probably taking a position once in the year and then keeping the position.

To reduce the statistical uncertainty unavoidable associated with events occurring rarely, we consider a smaller but large subset of active investors in the rest of the paper. Specifically, we consider only those investors who have 
traded the Nokia stock 
at least 20 days during the investigated time period. This means that all the investors in the subset have done at least 20 transactions, but not that all the investors who have participated to at least 20 transactions are in the set. The number of  investors fulfilling this requirement is 11,385 and they are responsible for 99.83\% of the volume exchanged during the considered period of time. The right part of Table~\ref{TableDescr1} gives  the number of investors, the number of transactions, and the traded volume for the restricted subset of investors.

Since the restricted set has a very high degree of heterogeneity in activity and in the characteristics of the investor, a key challenge is to devise methods allowing to compare and model trading actions performed by investors.
In this paper we use a recently developed statistical method based on network theory allowing to characterize clusters (or by using a technical term of network theory, communities) of heterogeneous investors \cite{Tumminello2011}.

\section{Categorical variables characterizing the trading activity}\label{categorical}

One of the consequences of the high degree of heterogeneity of investors is that it might be difficult to compare, for example, the activity of an household trading small volumes once every three months with a financial institution that trades every day large volumes. Moreover typical trading volume can be different in different periods of time, especially in the five year investigated period. Since we are interested in comparing the trading position taken by an investor in a given day, irrespective of the absolute volume traded, we introduce a categorical variable that describes its trading activity.  Specifically, for each investor $i$ and each trading day $t$, we consider the Nokia volume sold $V_s(i,t)$ and the Nokia volume purchased $V_b(i,t)$ by the investor in that day. This information is then converted into a categorical variable with 3 states: primarily buying \emph{b}, primarily selling \emph{s}, buying and selling with closing the position \emph{bs}.  The conversion is done by using the ratio
\begin{equation}
r(i,t)=\frac{V_b(i,t)-V_s(i,t)}{V_b(i,t)+V_s(i,t)}.
\end{equation}
We assign an investor a primarily buying state \emph{b} when $r(i,t)>\theta$,  a primarily selling state \emph{s} when $r(i,t)<-\theta$, and a buying and selling state \emph{bs} when $ -\theta \le r(i,t) \le \theta$  with $V_b(i,t)>0$ and $V_s(i,t)>0$. 
We have investigated the system by ranging $\theta$ between 0.01 and 0.25. The obtained results are not strongly affected by the specific choice of the threshold. In the present study we report results obtained by setting $\theta=0.01$. 

Given this categorization, we can map our system in a bipartite network where one set of nodes is composed by the investors and the other set by the trading days in the investigated data. A buy link is present between a node $i$ representing an investor and  a node $t$ representing a day if investor $i$ was in buying state \emph{b} at day $t$. Similarly we can define sell and buy-sell links. The bipartite network has therefore three different types of links. 

Here we are interested in the identification of clusters of investors. For this reason from the bipartite system we construct the projected network of investors. Note that, because there are three different types of links in the original bipartite system, there will be nine possible types of links in the projected network of investors, corresponding to different combinations of the actions of the two investors. 
The projected network is therefore composed by nodes representing investors and each pair of nodes can be linked by up to nine different links. Each link has a weight corresponding to the number of days in which the two investors are found in the pair of states characterizing the link.  
The projected network is almost a complete network and, more important, we want to preserve the information on the type of links joining the two nodes. In order to identify clusters in this complicated system, we preliminary identify those links that are statistically validated against a suitable null hypothesis. We perform this identification by using a recently introduced method \cite{Tumminello2011}, termed statistically validated networks. This
method has been demonstrated to be effective to investigate financial
and biological systems \cite{Tumminello2011}, as well as social
systems \cite{Tumminello2011b}. Next section explains how we apply this method to the investigated system.

\section{Statistically validated co-occurrence networks}\label{cooc}

\subsection{Method}

In the present investigation we use a statistical validation method of the co-occurrence of trading actions among heterogeneous investors coded by categorical variables. A co-occurrence is quantified by the presence of a weighted link of a specific type between two nodes (investors) in the projected network. Our method is robust with respect to the heterogeneity of trading activity of investors whereas we neglect the limited heterogeneity of state occurrence in different trading days because the number of \emph{b},  \emph{s} and \emph{bs}  states is only moderately fluctuating across different days and it has a bell shaped distribution with a range of fluctuations smaller than one decade. 

Within this approximation we can statistically validate the co-occurrence of state $P$ (either \emph{b},  \emph{s} or \emph{bs}) of investor $i$ and state $Q$ (either \emph{b},  \emph{s} or \emph{bs})  of investor $j$ with the following procedure. First of all, for each investor we identify its activity period, i.e. the time period in the investigated years when the investor was the owner of at least one share of any Finnish asset (not necessarily Nokia). Then for each pair $i$ and $j$ of investors we focus our attention on the intersection of the corresponding activity periods. We call $T$ the length of this intersection. 
In the intersection period of traders' activity, let us call $N_P$ ($N_Q$) the number of days when investor $i$ ($j$) is in the state $P$ ($Q$) and denote by $N_{P,Q}$ the number of days when we observe the co-occurrence of state $P$ for investor $i$ and state $Q$ for investor $j$. Under the null hypothesis of random co-occurrence of state $P$ for investor $i$ and state $Q$ for investor $j$, the probability of observing $X$ co-occurrences of the investigated states of the two investors in $T$ observations is described by the hypergeometric distribution, $H(X|T,N_P,N_Q)$ \cite{Tumminello2011}. We can therefore associate  a p-value with each pair of investors for each combination of the investigated states. 
Specifically, for each kind of co-occurrence of states $P$ and $Q$, the p-value is
\begin{equation}
\label{pval}
p(N_{P,Q})=1-\sum_{X=0}^{N_{P,Q}-1}H(X|T,N_P,N_Q).
\end{equation}
We indicate the states \emph{b}, \emph{s} and   \emph{bs} of investor $i$ as $i_{b}$, $i_{s}$  and $i_{bs}$ respectively. The nine possible combinations of the three trading states between investor $i$ and $j$ are ($i_{b}$,$j_{b}$), ($i_{b}$,$j_{s}$), ($i_{b}$,$j_{bs}$),
($i_{s}$,$j_{b}$), ($i_{s}$,$j_{s}$), ($i_{s}$,$j_{bs}$),
($i_{bs}$,$j_{b}$), ($i_{bs}$,$j_{s}$), 
and ($i_{bs}$,$j_{bs}$).

To
statistically validate the co-occurrence $N_{P,Q}$, the p-value
$p(N_{P,Q})$ must be compared with a statistical threshold $p$. One
might be tempted to simply set p=0.01 or p=0.05. However the
statistical validation of all nine possible co-occurrences of categorical states between all pairs of investors of our set is a multiple hypothesis test and therefore it needs a multiple hypothesis test correction \cite{Miller1981}.
Widely used multiple hypothesis test corrections are the Bonferroni and the False Discovery Rate (FDR) methods. The Bonferroni correction is in the present case $p_b=2 p_t/9(N_i (N_i-1))$ where $p_t$ is the chosen statistical threshold for the single test (in our case we choose $p_t=0.01$), the denominator of the correction is the number of considered investor pairs ($N_i (N_i-1)/2$) times  9, which is the number of different co-occurrences investigated for each pair of investors. A less stringent correction is the FDR \cite{Benjamini1995}, which is calculated as follows: p-values from all the
different k tests  ($k= 9 N_i (N_i-1)$ in the present case) are first
arranged in increasing order ($p_1<p_2<...<p_k$), and then FDR threshold is obtained by finding the largest $k_{max}$ such that $p_{k_{max}}<k_{max} ~ p_b$. In Ref.~\cite{Tumminello2011}, we called the projected network of elements' co-occurrences Bonferroni network when the correction used is the Bonferroni correction and FDR network when the correction used is the FDR correction. It is worth noting that the Bonferroni network is obtained under more restrictive statistical assumptions than the FDR network. 

\subsection{Results}

In the present study our aim is to construct the Bonferroni and the FDR network of investors trading the Nokia stock during the period 1998-2003.
Each pair of investors is characterized by the specific set of the above nine co-occurrences which are statistically validated. 
A validation of a specific co-occurrence is observed when the associated p-value is below the selected statistical threshold determined according to the selected multiple hypothesis testing correction.
We call a set of validations of co-occurrences as a {\it co-occurrence combination}. There are $2^9=512$ possible co-occurrence combinations. This number is certainly large and may suggest that 
attaining a parsimonious description of the system based on co-occurrence combinations can be unlikely. However, we find that the number of observed distinct co-occurrence combinations is 19 for the Bonferroni network and 74 for the FDR network. It is worth noting that $99\%$ of relationships observed among investors in the Bonferroni and FDR network are described by just 6 and 9 distinct co-occurrence combinations, respectively.

These co-occurrence combinations are listed in Table~\ref{co_com}.
The co-occurrence combination C1 indicates that investors $i$ and $j$ 
show a validated co-occurrence of the trading action of primarily buying. Similarly, C2, C3, C5, C6 and C7 indicate various co-occurrences of the three considered actions. In our setting, co-occurrence combinations are not directional and therefore in each of the considered combinations the label $i$ and $j$ can be interchanged. Co-occurrence combination C4 presents a twofold validation involving the co-occurrences  ($i_{b}$,$j_{b}$) and ($i_{s}$,$j_{s}$). When this co-occurrence combination is observed, the two investors act synchronously both when they decide to buy and when they decide to sell.
C7 is the only 
significantly populated co-occurrence combination in which the two agents systematically take opposite trading position, i.e. one agent is buying when the other one is selling.
Our analysis shows that this kind of co-occurrence combination is only marginally probable (only 0.81 percent of the cases) 
when the multiple test correction is not too severe -- as it is the case for the FDR network. The co-occurrence combination C8 describes co-occurrence of all three states (primarily buying, primarily selling, and buying and selling) indicating a very strong level of synchronization between distinct investors. The two-fold co-occurrence combination C9 in turn describes co-occurrence of buying, and buying and selling activity together with selling, and buying and selling activity. This kind of relationship, for example, can be interpreted as describing the interaction between an investor and a market maker (or a day trader) acting coherently during the same days.

\begin{table}[t]
\begin{center}
\caption{Most populated co-occurrence combinations in the Bonferroni and FDR networks. The third (forth) column gives the number of the corresponding links in the Bonferroni (FDR) network. The number in the parenthesis gives the percentage of the corresponding links over the total number of links. The fifth column is the color label used for the links in Figure \ref{BonfClu} and \ref{NetF8}.  }
\vskip 0.5cm
\begin{tabular}{|l|l|c|c|c|}
\hline
Label & Co-occurrence & Bonferroni &  FDR  & Color\\
~ & combination & 36,664 links & 330,404 links & label \\
\hline
\hline
C1 & ($i_{b}$,$j_{b}$) & 7,716 (21.0) & 120,655 (36.5) & magenta \\
\hline
C2 & ($i_{s}$,$j_{s}$) & 6,254 (17.1) & 91,219 (27.6) & green \\
\hline
C3 & ($i_{bs}$,$j_{bs}$) & 1,732 (4.72) & 19,227 (5.82) & apricot \\
\hline
C4 & ($i_{b}$,$j_{b}$) & 20,243 (55.2) & 66,692 (20.2) & black \\
~ & ($i_{s}$,$j_{s}$) & ~ & ~ & ~ \\
\hline
C5 & ($i_{b}$,$j_{bs}$) & 312 (0.85) & 13,494 (4.08) & blue \\
\hline
C6 & ($i_{s}$,$j_{bs}$) & 157 (0.43) & 9,592 (2.90) & orange \\
\hline
C7 & ($i_{s}$,$j_{b}$) & 12 (0.033)  & 2,662 (0.81) & tan \\
\hline
C8 & ($i_{b}$,$j_{b}$) & 137 (0.37)  & 2,304 (0.70) & brown \\
~ & ($i_{s}$,$j_{s}$) & ~ & ~ & ~ \\
~ & ($i_{bs}$,$j_{bs}$) & ~ & ~ & ~ \\
\hline
C9 & ($i_{b}$,$j_{bs}$) & 43 (0.12)  & 1,414 (0.43) & purple\\
~ & ($i_{s}$,$j_{bs}$) & ~ & ~ & ~ \\
\hline
\hline
\end{tabular}
\label{co_com}
\end{center}
\end{table}

In summary, the different investors are connected in the Bonferroni and in the FDR networks by links of different nature each of them describing a specific co-occurrence combination. This structure is richer than a customary unweighted network, and it is also different from a weighted network because co-occurrence combinations describes relationships which cannot be described by a numerical value only. In Ref.~\cite{Tumminello2011}, we addressed this kind of links present in a statistically validated network with the term \emph{multi-link}.

 \begin{table*}[ht]
\begin{center}
  \caption{Summary of the number of investors (\# ids), the number of transactions ($N$), and the exchanged volume ($V$, in millions of shares) for Nokia investors included in the Bonferroni network and in the FDR network. The investors are divided in the six categories (top part) or between nominee and non nominee registered.}
\begin{tabular}{l|rrr|rrr}
 \hline
  ~	&	~	& Bonferroni & ~ & ~ & FDR & ~ \\
  \hline
  Category				& \# ids&$N$~~~ 	&$V$~~ 	&\#ids	&$N$~~~	&$V$~~	\\
  \hline
 Non financial corporations 	&580	&	 1,202,142& 6,847		&1,472	&1,410,377	&8,637	\\
 Financial and insurance 		&112	&	 7,316,946& 64,344		&185	&14,039,035	&171,145	\\ 
 Governamental			&61		&	        15,223& 222		&75		&20,462		&615	\\
 Non profit					&53		&	        10,684&79			&95		&17,470		&261	\\
 Households				&2,292	&	      501,620&595		&8,521	&968,268 		&903	\\
 Foreign organizations		&20		&	        29,933&36			&87		&330,580		&2,482	\\
 \hline
Total						&3,118	&9,076,548	&72,123		&10,435	&16,786,192	&184,043	\\
\hline
\hline
Nominee registered			& 18 		& 6,096,148 & 61,060		& 31		&12,386,315	&166,353 \\
Non nominee registered		& 3,100	& 2,980,400 & 11,063		&10,404	&4,399,877	&17,690 \\
\hline
\end{tabular}
  \label{TableDescr2}
\end{center}
\end{table*}

Table \ref{TableDescr2} gives the number of investors, the number of transactions, and the traded volume for the six categories for the subset of investors in the Bonferroni (left part) and in the FDR (right part) network. It also gives these numbers separating nominee registered and non nominee registered investors. The Bonferroni network of investors is composed by $3,118$ investors connected by 36,664 multi-links.  
The Bonferroni network presents 226 connected components. The largest one has 2,537 investors and the other connected components are much smaller, with size ranging from 26 to 2 investors. The FDR network covers almost completely the set of investigated investors. In fact, it is composed by 10,435 investors connected by 330,404 multi-links. 
The FDR network includes 22 connected components. The largest one has 10,389 investors (99.6\% of the network's investors) while the other connected components are extremely small with a size of only 3 or 2 investors.

\section{Cluster detection in co-occurrence networks}\label{community}

We are now interested in finding a partition of the Bonferroni and FDR networks that reveals the cluster structure of investors. The largest clusters detected both in the Bonferroni and in the FDR networks shall be analyzed with methods of community detection in network to reveal the clusters of investors characterized by similar trading profile. 
As discussed in the previous section, multi-link statistically validated networks of investors are networks presenting different
classes of links. To the best of our knowledge, there is no established method specifically devised to detect communities in
networks with links of qualitative different nature, such as our statistically validated networks.
Here we propose a minimalist approach by removing from the network all the co-occurrence combinations which validate opposite trading actions of the considered pair of investors (as, for example, the C7 co-occurrence combination of Table \ref{co_com}). From the cited Table we see that this is a very limited set of links covering 0.81\% of the links in the FDR and 0.033\% in the Bonferroni network. 
Moreover it is reasonable to expect that investors belonging to the same cluster display similar trading behavior and therefore, these anti-diagonal links are likely to bridge different clusters of investors.

We perform community detection in the modified Bonferroni and FDR networks by using the Infomap method  \cite{Rosvall2008}. This method is considered as one of the most effective methods of community detection in networks \cite{Lancichinetti2009}. We apply the method by considering weighted networks where the weight of each multi-link is given accordingly to the number of co-occurrence validations observed in the co-occurrence combination. 
For example, the weight of C1 combination ($i_b,j_b$)  is 1 and the weight of C8 combination (($i_b,j_b; i_s,j_s;i_{bs},j_{bs}$) is 3.
While our approach is pragmatic and heuristic, we are aware that  a more theoretically grounded approach to partitioning multi-link networks would certainly be useful in the study of networks where links of different nature can be naturally defined, as in the present case.

\subsection{Clusters in the Bonferroni network}

\begin{figure}[t]
\begin{center}
\includegraphics[scale=0.36]{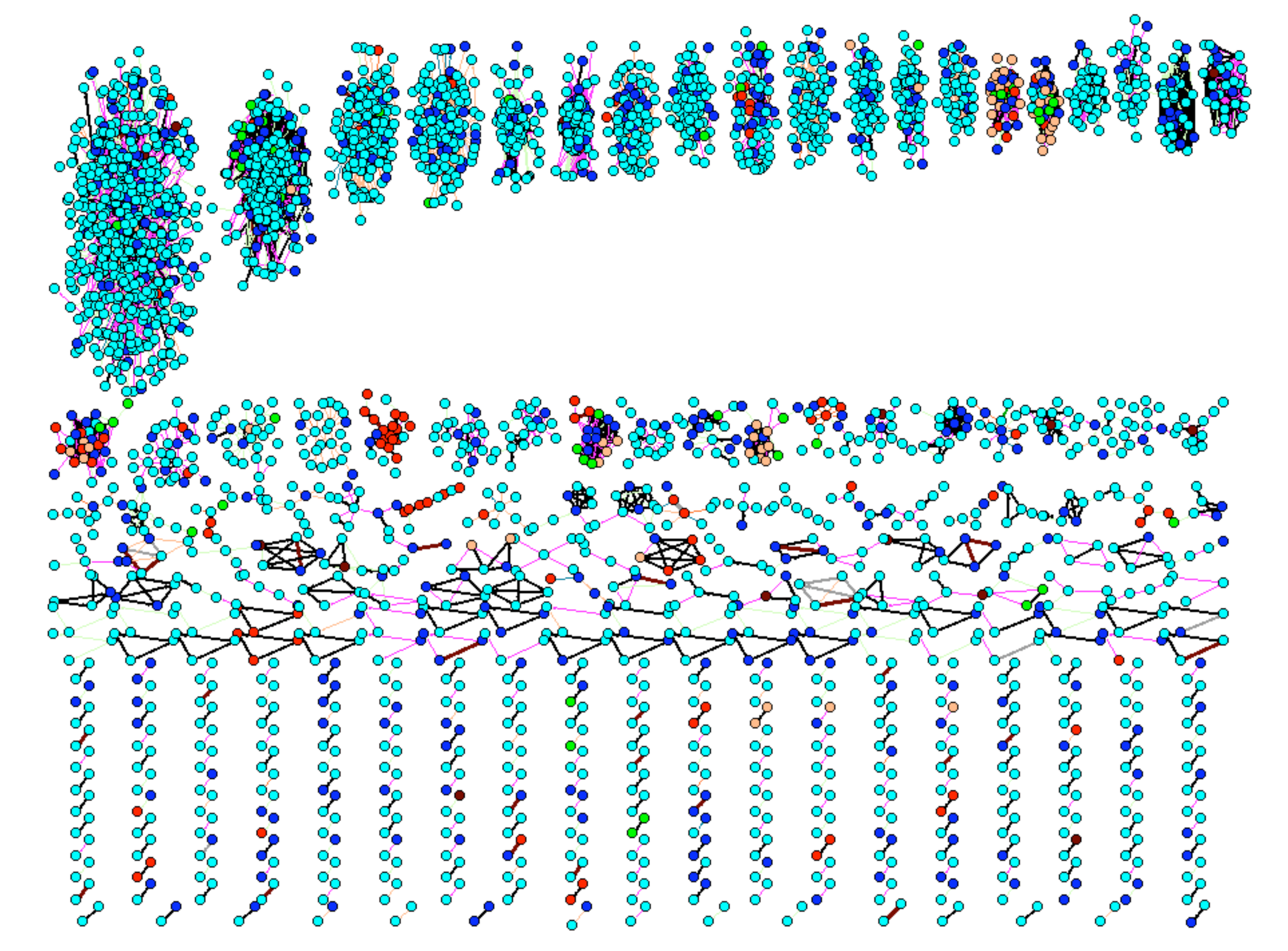}

\includegraphics[scale=0.36]{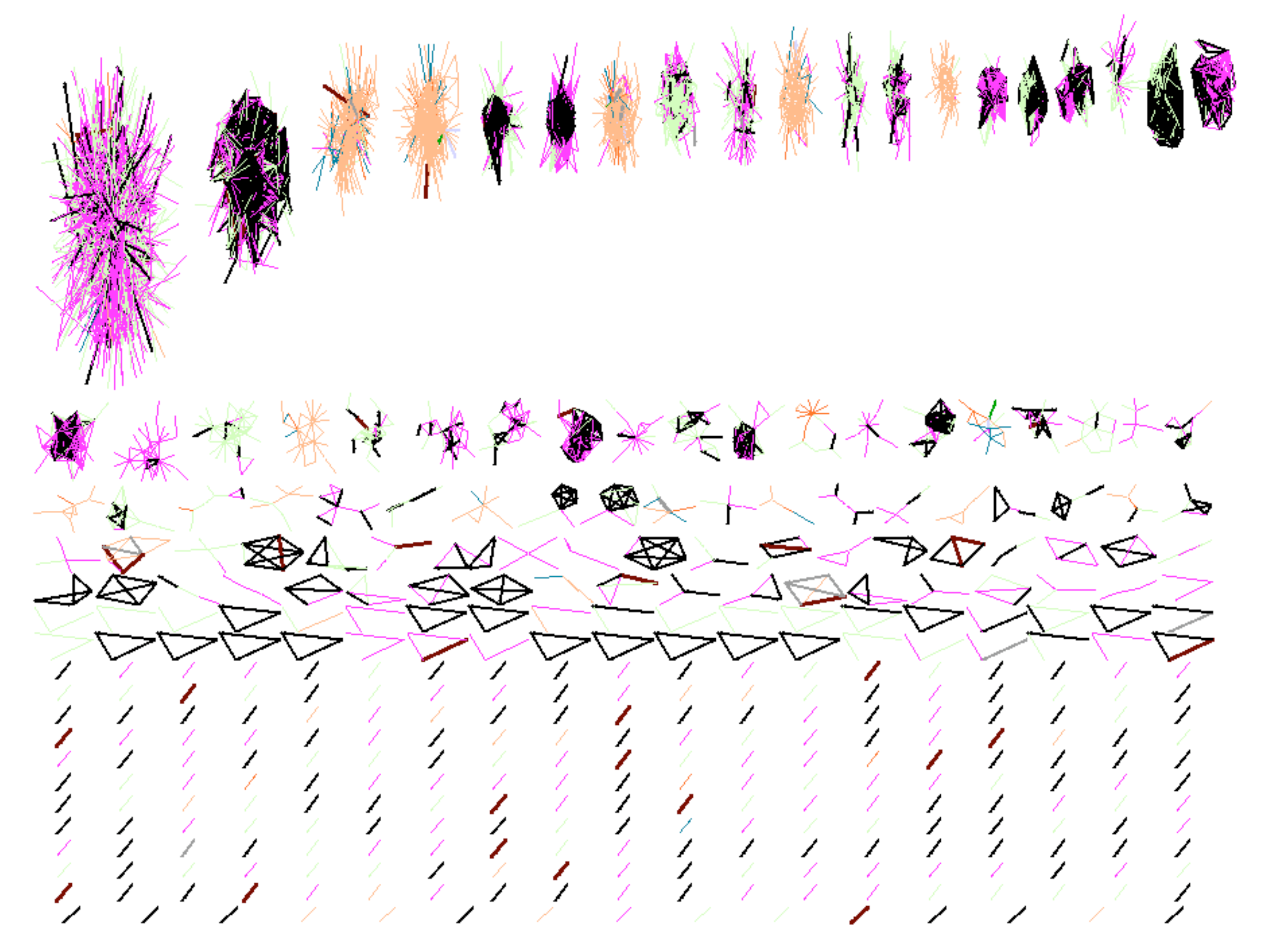}
\caption{Clusters of investors detected by the Infomap method in the Bonferroni SVN. Top panel: Emphasis on the vertices of the clusters. The color of the vertex indicates the category that the investor belongs to.  The colors are assigned as follows: corporations (blue), general governmental organizations (yellow), foreign organizations (maroon), non-profit institutions (green),  financial and insurance corporations (red) and households (cyan). Bottom panel: Same clusters as in the top panel but in this case we remove vertices and emphasise the nature of the links connecting investors. The color code of the links is provided in Table~\ref{co_com}.}
\end{center}
\label{BonfClu}
\end{figure}

In Fig.~\ref{BonfClu} we show the clusters detected by the Infomap algorithm in the Bonferroni network. We observe 356 distinct clusters of size ranging from 527 to 2. The largest 10 clusters comprise 1,522 investors, approximately half the number of investors in the Bonferroni SVN. In Fig.~\ref{BonfClu} we plot all the detected clusters in two panels. The top panel displays each investor in the network with a circle whose color indicates the category to whom the investor belongs to. The most common investors are households (cyan circles) which are also the most numerous investors in the database. However, in some cases an over-representation of some other category of investors (for example, financial corporations labeled as red circles) is apparent. Below we will provide rigorous results about the over-representation or under-representation of categories of investors in the detected clusters.  
In the bottom panel of the figure, nodes are not colored and their size is reduced to point out the nature of the links connecting the nodes in different clusters. The color code of different links is given in Table~\ref{co_com}. The figure shows that several clusters are characterized by a specific (single color) multi-link. 

\begin{figure}[Ht]
\begin{center}
\includegraphics[scale=0.45]{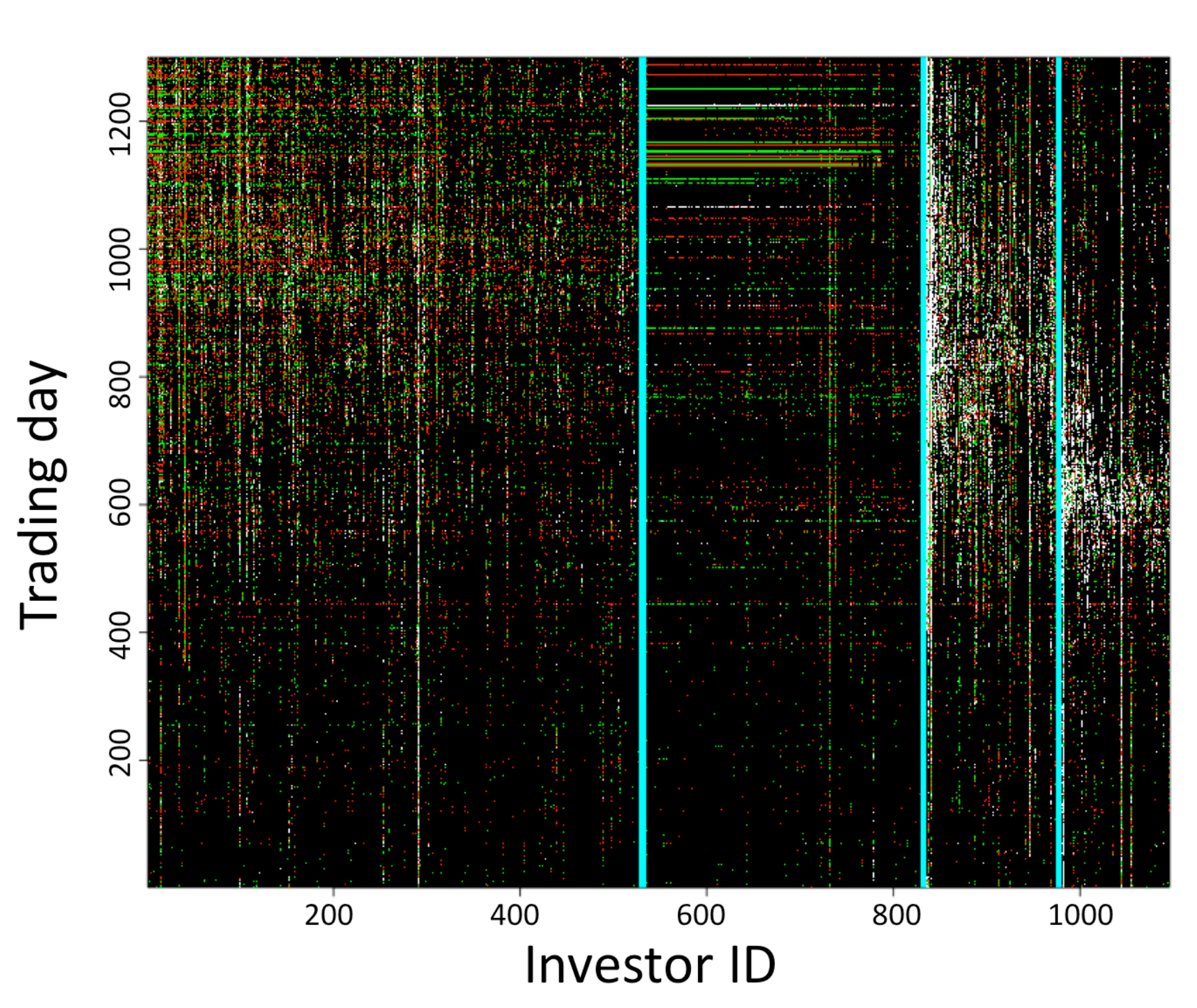}

\includegraphics[scale=0.45]{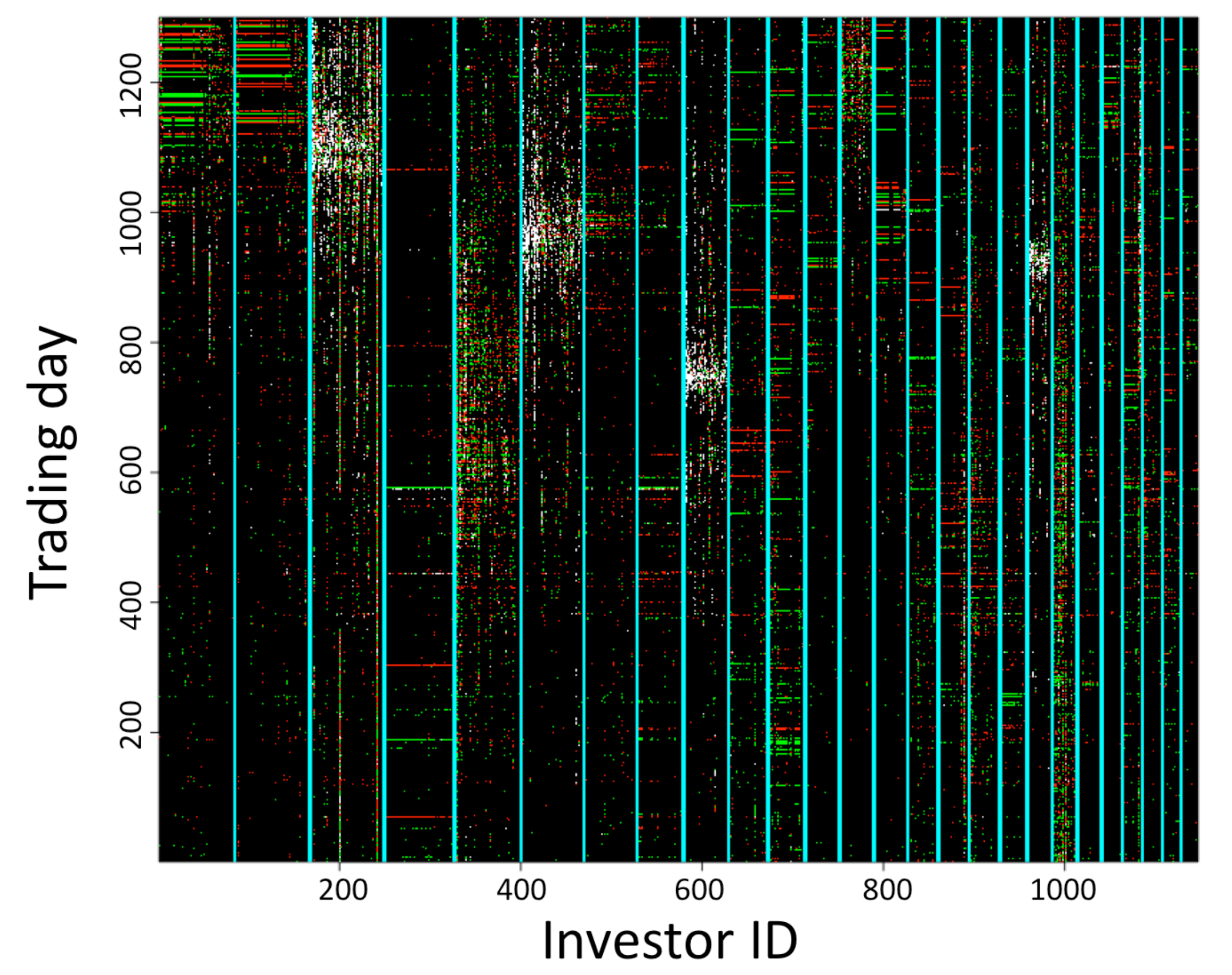}
\caption{Microarray-like representation of the trading activity of investors of the 30 most populated clusters detected in the Bonferroni statistically validated network by the Infomap method.  A red spot indicates a buying trading action of an investor, a green spot a selling action, and a white spot a buying and selling trading action. Different investors are ordered along the $x$-axis. The $y$-axis is time (in trading days). In the absence of trading activity the corresponding spot is black. Top panel: the top 4 clusters (B1 to B4 from left to right). Bottom panel: the remaining 26 clusters (B5 to B30 from left to right). Vertical light blue lines separate the clusters.}
\end{center}
\label{ArrayBon30}
\end{figure}

\begin{table*}
\begin{center}
\caption{Summary statistics of the 30 most populated clusters of the Bonferroni network detected with Infomap. For each cluster we statistically validate the over-expression or under-expression of investors belonging to a specific category: non-financial corporations  (C), general governmental organizations (G), foreign organizations (FO), non-profit institutions (NP), financial and insurance corporations (FI), and households (H). We also statistically validate the over-expression or under-expression of multi-links belonging to a specific co-occurrence combination. The list of most frequent co-occurrence combinations are given in Table \ref{co_com}.}
\vskip 0.5cm
\begin{tabular}{|l||r|c|c|c|c|} \hline
Cluster & Investors & Over-expr. & Under-expr. & Over-expr. & Under-expr. \\
~ & ~ & investor category & Investor category & co-occur. comb. & co-occur. comb. \\
\hline
B1 & 527 & H & C G  NP FI & C1 C2 & C3 C4 \\
B2 & 294 & ~ & FI & C4 & C1 C2 C3 C5 C6  C9 \\ 
B3 & 138 & ~ & ~ & C3 C5 C6 C9 & C1 C2 C4 \\ 
B4 & 116 & ~ & ~ & C3 & C1 C2 C4 \\ 
B5 & 82 & ~ & ~ & C4 & C1 C2 C3 \\ 
B6 & 79 & ~ & ~ & C1 C4 C5 & C2 C3 C8 \\ 
B7 & 78 & ~ & ~ & C3 C5 C6 C9 & C1 C2 C4 \\ 
B8 & 73 & ~ & ~ & C2 & C1 C3 C4 \\ 
B9 & 70 & ~ & ~ & C1 C2 & C4 \\ 
B10 & 65 & ~ & ~ & C3 C5 & C1 C2 C4 \\ 
B11 & 55 & ~ & ~ & C2 & C1 C3 C4 \\ 
B12 & 47 & ~ & ~ & C1 C2 & C3 C4 \\ 
B13 & 46 & ~ & ~ & C3 & C1 C2 C4 \\ 
B14 & 39 & G NP & H & C1 & C2 C3 C4 \\ 
B15 & 37 & G NP & H & C2 & C1 C3 \\ 
B16 & 34 & ~ & ~ & ~ & C3 \\ 
B17 & 34 & ~ & ~ & C1 C2 & C4 \\ 
B18 & 33 & ~ & ~ & C4 & C1 C3 \\ 
B19 & 30 & FI & H & C1 & C3 C4 \\ 
B20 & 30 & ~ & ~ & C1 & C2 C3 C4 \\ 
B21 & 30 & ~ & ~ & C1 & C2 C4 \\ 
B22 & 26 & ~ & ~ & C2 & C4 \\ 
B23 & 24 & ~ & ~ & C3 & C4 \\ 
B24 & 23 & FI & H & C2 & C4 \\ 
B25 & 23 & ~ & ~ & C1 & C4 \\ 
B26 & 19 & ~ & ~ & C1 & C4 \\ 
B27 & 18 & G NP & H & C1 & C2 \\ 
B28 & 18 & ~ & ~ & C1 & C4 \\ 
B29 & 17 & G & H & ~ & ~ \\ 
B30 & 17 & ~ & ~ & C2 & C4 \\ 
\hline
\hline
\end{tabular}
\label{Bonf30}
\end{center}
\end{table*}

 In Fig.~\ref{ArrayBon30} we show the trading activity of the 30 most populated clusters by using a representation where a red spot corresponds to the buying action of an investor, a green spot to the selling action and a white spot to the buying and selling action. In the absence of trading activity the corresponding spot is black. Different investors are ordered along the $x$-axis and we put time in the $y$-axis (in trading days). For the sake of readability we separate clusters by light blue vertical lines. Before investigating the composition of these clusters and the over-representation of specific categories, it is worth emphasizing a clear result emerging from Fig.~\ref{ArrayBon30}. Most clusters are characterized by a very high degree of synchronization in the timing of the trading activity among the investors of the cluster. In other words, for many clusters we observe that a large fraction of the investors in the cluster trades in the same days (and often with the same trading state). 
 
We then analyze the clusters of investors detected in the Bonferroni network by using the information available about the category of investors which are present in each cluster. A statistical method to perform an analysis of over-representation of attributes of elements partitioned in clusters is given in Ref.~\cite{Jstat}. This method is needed to provide a statistical validation of over-expression and under-expression because the categories of investors and the co-occurrence combinations are quite heterogeneous in number, and this aspect needs to be taken into account properly in the analysis \cite{Jstat}. 
In Table \ref{Bonf30} we present the summary information about the 30 most populated clusters detected with the Infomap method in the Bonferroni SVN. Specifically, for each cluster we indicate the number of investors, the over-expression or under-expression of category of investors, and the over-expression or under-expression of multi-links belonging to a co-occurrence combination.  

A description of major properties of clusters can be achieved by analyzing jointly Fig.~\ref{ArrayBon30} and the information summarized in Table \ref{Bonf30}. As expected, from the Figure we note that the trading activity is quite heterogeneous for investors of different clusters. In some clusters, for some periods of time, the trading activity is rather continuous whereas other clusters show a sparse trading activity localized around to specific days. Fig.~\ref{ArrayBon30} also shows that there are clusters of investors (most probably so-called market makers and day traders, i.e. investors closing their position without a net inventory at the end of the trading day) characterized by a buying and selling intraday activity (clusters B3, B4, B7, B10, B13 and B23). Within the clusters investors interact among them and with other buying and selling investors in rather localized period of time. In fact, in Table \ref{Bonf30} we note that the over-expressed multi-links are of kind C3 ($i_{bs}$,$j_{bs}$), C5 ($i_{b}$,$j_{bs}$), C6 ($i_{s}$,$j_{bs}$) and C9 ($i_{b}$,$j_{bs}$) \& ($i_{s}$,$j_{bs}$). It is worth noting that for these clusters no over-expression or under-expression of category of investors is detected suggesting that trading activities requiring a daily closure of the trading position is present in investors of all categories with a degree of heterogeneity similar to the one of the entire set of investors.

The other clusters present trading activities characterized by the co-occurrence of buying, selling and/or buying and selling trading actions. Clusters characterized by an over-expression of multi-link C4 ($i_{b}$,$j_{b}$) \& ($i_{s}$,$j_{s}$) are clusters B2, B5, B6 and B18. These clusters present the highest degree of synchronicity among investors both when they
 decide to buy and when they decide to sell. The remaining clusters present an over-expression of buying co-occurrence C1 ($i_{b}$,$j_{b}$) (specifically, clusters B14, B19, B20, B21, B25, B26, B27, B28) or of selling co-occurrence C2 ($i_{s}$,$j_{s}$)   (specifically, clusters B8, B11, B15, B22, B24, B30). 
Finally clusters B1, B9, B12, and B17 present over-expressions of C1 ($i_{b}$,$j_{b}$) and C2 ($i_{s}$,$j_{s}$) multi-links. It should be noted that this case is different from the over-expression of C4 ($i_{b}$,$j_{b}$) \& ($i_{s}$,$j_{s}$) because the C1 and C2 over-expression can involve different pairs of investors whereas the C4 over-expression involves the same pairs of investors. In other words, when C1 and C2 co-occurrence combinations are observed separately, this observation reflects the fact that subsets of the investors are coherently buying among them and other subsets are coherently selling with a non null intersection among the subsets, i.e. an investor can coherently buy with another one whereas it is coherently selling with a third one and so on.  

Some clusters (B1, B14, B15, B19, B24, B27, and B29) present over-expression and under-expression of investors belonging to specific categories. Cluster B1 presents an over-expression of households and an under-expression of non-financial corporations, general governmental organizations, non-profit institutions and financial and insurance corporations. The trading strategies of the underlying cluster are therefore those ones which are most popular among single individuals. Clusters B14, B15 and B27 present an over-expression of general governmental organizations and non-profit institutions and an under-expression of Households. The trading strategies underlying these clusters are therefore trading strategies which must be popular among this kind of institutions (Governmental and non profit).
Finally, clusters B19, and B24 present an over-expression of financial and insurance corporations and an under-expression of households.

\subsection{FDR network and its relation with the Bonferroni network}
 
We have also computed the FDR network of the investors. As expected, it includes more investors ($10,435$) and more multi-links (330,404) than the Bonferroni network, since the requirement on the statistical validation is less restrictive. The FDR network has a large  connected component comprising $10,389$ investors. By applying the Infomap algorithm to the FDR network where we remove co-occurrence combinations of the kind
($i_{b}$,$j_{s}$) and we obtain a partition of investors in 390 clusters whose size is ranging from 3,000 to 2 investors. The 30 most populated clusters are described in Table \ref{FDR30} where we report the over-expression and under-expression of categories of investors and co-occurrence combinations observed for each cluster.  

\begin{table*}
\begin{center}
\caption{Summary statistics of the 30 most populated clusters of the FDR network detected with Infomap. For each cluster we statistically validate the over-expression or under-expression of investors belonging to a specific category: non-financial corporations  (C), general governmental organizations (G), foreign organizations (FO), non-profit institutions (NP), financial and insurance corporations (FI) and households (H). We also statistically validate the over-expression or under-expression of multi-links belonging to a specific co-occurrence combination. The list of most frequent co-occurrence combinations are given in Table \ref{co_com}.}
\vskip 0.5cm
\begin{small}
\begin{tabular}{|l||r|c|c|c|c|} \hline
Cluster & Investors & Over-expr. & Under-expr. & Over-expr. & Under-expr. \\
~ & ~ & investor category & Investor category & co-occur. comb. & co-occur. comb. \\
\hline
F1 & 3000 & H & G  NP FI & C1 C2 C5 C6 C9 & C4 C3 C8 \\
F2 & 1851 & H & C G & C1 & C2 C3 C4 C5 C6 C8 C9 \\ 
F3 & 931 & ~ & G & C3 C5 C6 C9 & C1 C2 C4 C8 \\ 
F4 & 639 & ~ & ~ & C1 C4 C9 & C2 C3 C5 C6 C8 \\ 
F5 & 438 & C NP & H & C4 C8 & C1 C2 C3 C5 C6 C9 \\ 
F6 & 312 & FI & ~ & C2 C5 C6 & C4 C8 \\ 
F7 & 223 & ~ & ~ & C3 C5 C6 & C1 C2 C4 \\ 
F8 & 205 & C  G FI NP & H & C4 & C1 C2 C3 C5 C6 C8 C9 \\ 
F9 & 140 & ~ & ~ & C3 C5 C6 C9 & C1 C2 C4 \\ 
F10 & 129 & ~ & ~ & C2 C4 & C1 C3 C5 C6 C9 \\ 
F11 & 127 & ~ & ~ & C3 C5 C6 C9 & C1 C2 C4 \\ 
F12 & 85 & ~ & ~ & C2 & C1 C3 C4 C5 C6 \\ 
F13 & 68 & ~ & ~ & C4 & C1 C3 C5 C6 \\ 
F14 & 54 & ~ & ~ & C3 C5 C6 & C1 C2 C4 \\ 
F15 & 40 & ~ & ~ & C4 & C2 C3 C5 \\ 
F16 & 39 & ~ & ~ & C4 & C1 C2 C3 C5 C6 \\ 
F17 & 39 & ~ & ~ & C4 & C2 C3 C5 C6 \\ 
F18 & 37 & ~ & ~ & C1 & ~ \\ 
F19 & 29 & ~ & ~ & C4 & C2 \\ 
F20 & 26 & ~ & ~ & C2 & C1 \\ 
F21 & 26 & ~ & ~ & C6 & C3 \\ 
F22 & 24 & ~ & ~ & C6 & ~ \\ 
F23 & 22 & ~ & ~ & C4 C8 & C1 \\ 
F24 & 20 & ~ & ~ & C8 & C2 \\ 
F25 & 19 & ~ & ~ & C4 & C1 \\ 
F26 & 19 & ~ & ~ & C2 & C1 \\ 
F27 & 17 & ~ & ~ & ~ & ~ \\ 
F28 & 16 & ~ & ~ & ~ & ~ \\ 
F29 & 16 & ~ & ~ & ~ & ~ \\ 
F30 & 16 & ~ & ~ & ~ & ~ \\ 
\hline
\hline
\end{tabular}
\label{FDR30}
\end{small}
\end{center}
\end{table*}

Clusters detected in the Bonferroni and in the FDR network are related. The most common relationship is inclusiveness of the Bonferroni clusters into the FDR clusters. In fact, 23 of the 30 largest Bonferroni clusters have more than 75\% of their elements in corresponding single FDR clusters. By requiring more than 90\% of the elements of a Bonferroni cluster to be in a single FDR cluster this number reduces to 17. In Table \ref{FDR_BON30} we show the inclusiveness relationships observed for the most populated Bonferroni clusters when more than 75\% of the elements are present in corresponding single FDR clusters.  

While the inclusiveness of Bonferroni clusters into FDR clusters is the most common relationship, we  also observe partitioning of the elements of a Bonferroni cluster into many FDR clusters. For example,
the elements of cluster B7 are sorted out into clusters F1 (40\%), F4 (14\%) and F11 (46\%). This is due to the fact that the inclusion of new multi-links in the FDR network sometimes can significantly change the local density of multi-links around specific regions of the network. These changes of the local multi-link structure can therefore be reflected into the partitioning performed by community detection algorithms.

Another aspect to be taken into account concerns the nature of links of statistical validated networks. In the the present case links are multi-links of different nature and the co-occurrence combination between two investors can be different  in the Bonferroni and in the FDR networks. For example, a multi-link between investors $i$ and $j$ can be of C1 type in the Bonferroni network and of C4 type in the FDR network due to the further statistical validation of the co-occurrence ($i_{s}$,$j_{s}$) when the FDR multiple test correction is used. 
The percent of Bonferroni multi-links which are changing nature when detected in the FDR network is close to 37\%. However, 30\% of them concerns the co-occurrence combinations C1 (16\%) and C2 (14\%). Both co-occurrence combinations change to the C4 co-occurrence combination validating both the co-occurrence of buying and selling. In other words, in some cases, the strictest Bonferroni correction validates only the buying or the selling co-occurrence, whereas with the FDR multiple test correction the co-occurrence is validated both for buying and selling.
\begin{table}
\begin{center}
\caption{Inclusiveness relationships of the 30 most populated Bonferroni clusters. The relationship is indicated  when more than 75\% of the elements of a Bonferroni cluster is present into a single FDR cluster. An asterisk indicates that more than 90\% of the elements are present in the corresponding FDR cluster.}
\vskip 0.5cm
\begin{tabular}{|l||c|} \hline
FDR Cluster & Bonferroni clusters\\
\hline
F1 & B1 (*) B10(*) B11 B23 (*) \\
F2 & B21 (*) \\
F3 & B4 B13 (*) \\
F4 & B5 B6 (*) B17 (*)\\
F5 & B2 (*) B26 (*)\\
F8 & B14 (*) B15 (*) B19 (*) B27 B29 (*)\\
F10 & B8 (*) \\
F12 & B22 (*) \\
F13 & B12 (*) \\
F15 & B25 (*) \\
F16 & B16 \\
F17 & B20 \\
\hline
\hline
\end{tabular}
\label{FDR_BON30}
\end{center}
\end{table}

\begin{figure}[t]
\begin{center}
\includegraphics[scale=0.34]{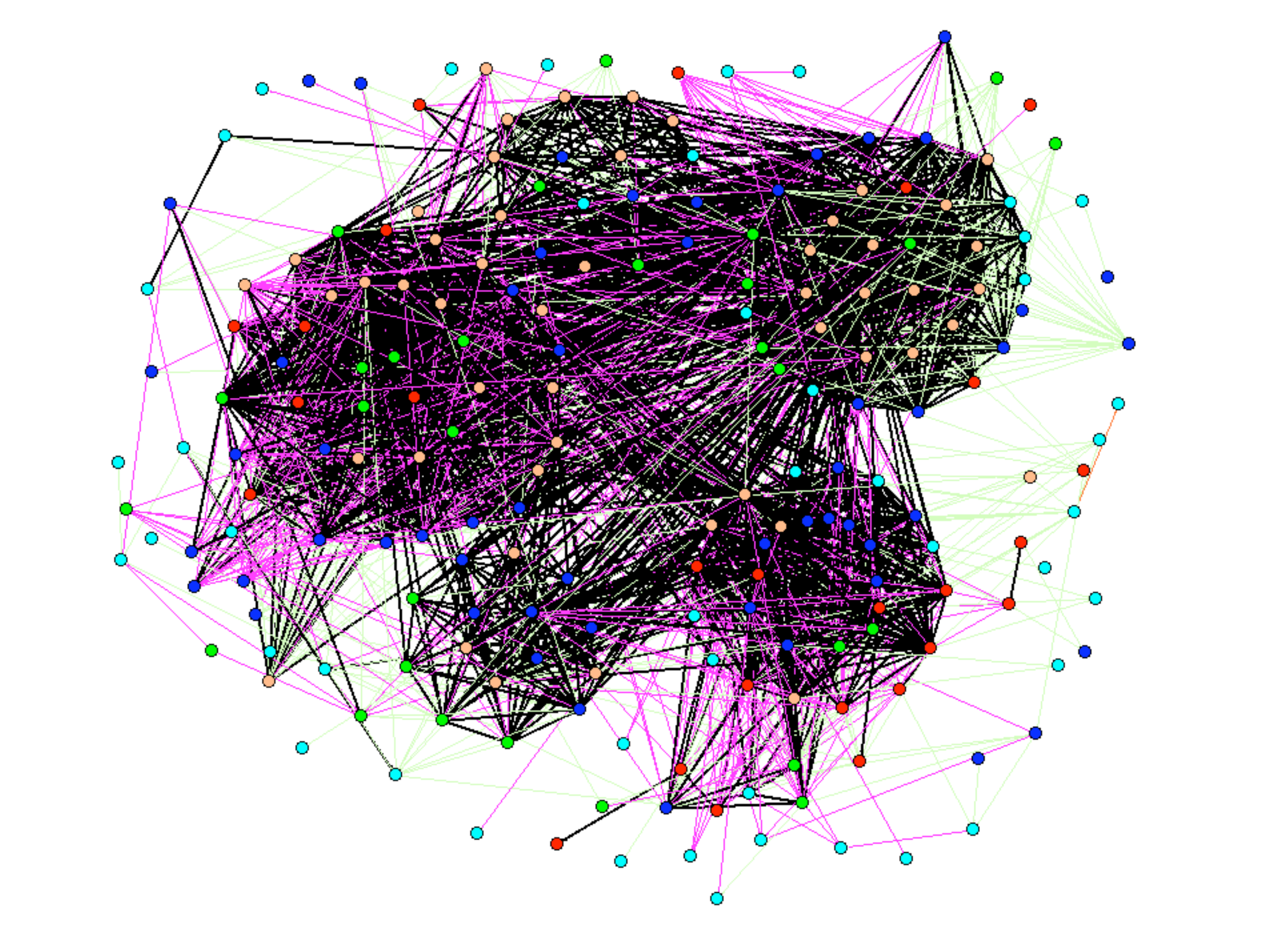}

\includegraphics[scale=0.34]{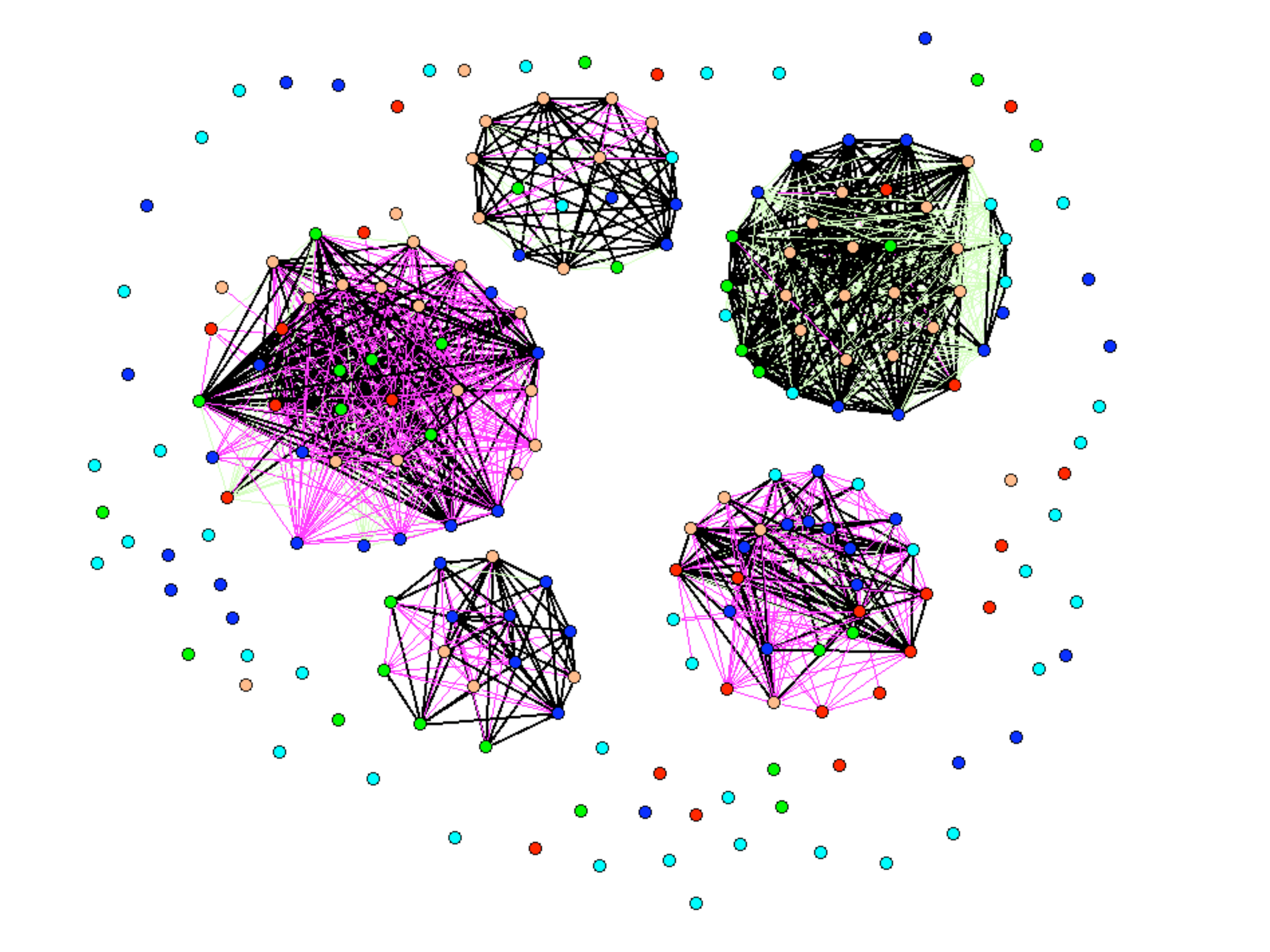}
\caption{Network of investors belonging to the F8 cluster of the FDR network (top panel) and to B14, B15, B19, B27 and B29 clusters of the Bonferroni network (bottom panel) that are included in the F8 FDR cluster. The color of vertices is given as indicated in the caption of Fig. \ref{BonfClu}. The color code of links is provided in Table \ref{co_com}. In the top panel we show the links of the FDR network whereas in the bottom panel we show the links of the Bonferroni network. The 5 clusters of the Bonferroni network are from top in clockwise order B29, B15, B19, B27 and B14. B15 shows a over-representation of 
C2 ($i_{s}$,$j_{s}$) links (green links), whereas B14, B19, and B27 clusters have C1 ($i_{b}$,$j_{b}$) (magenta links). In the F8 FDR cluster the over-represented link  is C4 ($i_{b}$,$j_{b}$) \& ($i_{s}$,$j_{s}$) (black links).}
\end{center}
\label{NetF8}
\end{figure}

We provide a concrete example of the above discussed concepts by comparing the F8 cluster of the FDR network and B14, B15, B19, B27 and B29 Bonferroni clusters included in it. In the top panel of Fig.~\ref{NetF8} we show the F8 cluster of the FDR network. From Table \ref{FDR30} we note that the F8 cluster has over-expression of investors belonging to the non-financial corporations, general governmental organizations, non-profit institutions, and financial and insurance corporations categories, whereas household investors are under-expressed. The same table shows that the over-expressed co-occurrence combination is the C4 combination (black links in Fig.~\ref{NetF8} ) implying the ($i_{b}$,$j_{b}$) and ($i_{s}$,$j_{s}$) co-occurrence. In the bottom panel of the same figure we show the five clusters B14, B15, B19, B27 and B29 of the Bonferroni network (we show only the 135 elements which are also present in the F8 cluster of the FDR). The   
links present in the bottom panel are the links of the Bonferroni network. By comparing the top panel (FDR cluster) and the bottom panel (Bonferroni clusters), we note that the Bonferroni clusters describe core regions of the wider FDR clusters, and that the number of multi-links grows if we move from the Bonferroni to the FDR network. In some cases, links change nature from C1 and C2 co-occurrence combinations to C4. In fact, by analyzing Table \ref{Bonf30}, we see that the over-expressed multi-links of the Bonferroni clusters are C1 for B14, B19, and B27, and  C2 for B15.

\begin{figure}
\begin{center}
\includegraphics[scale=0.45]{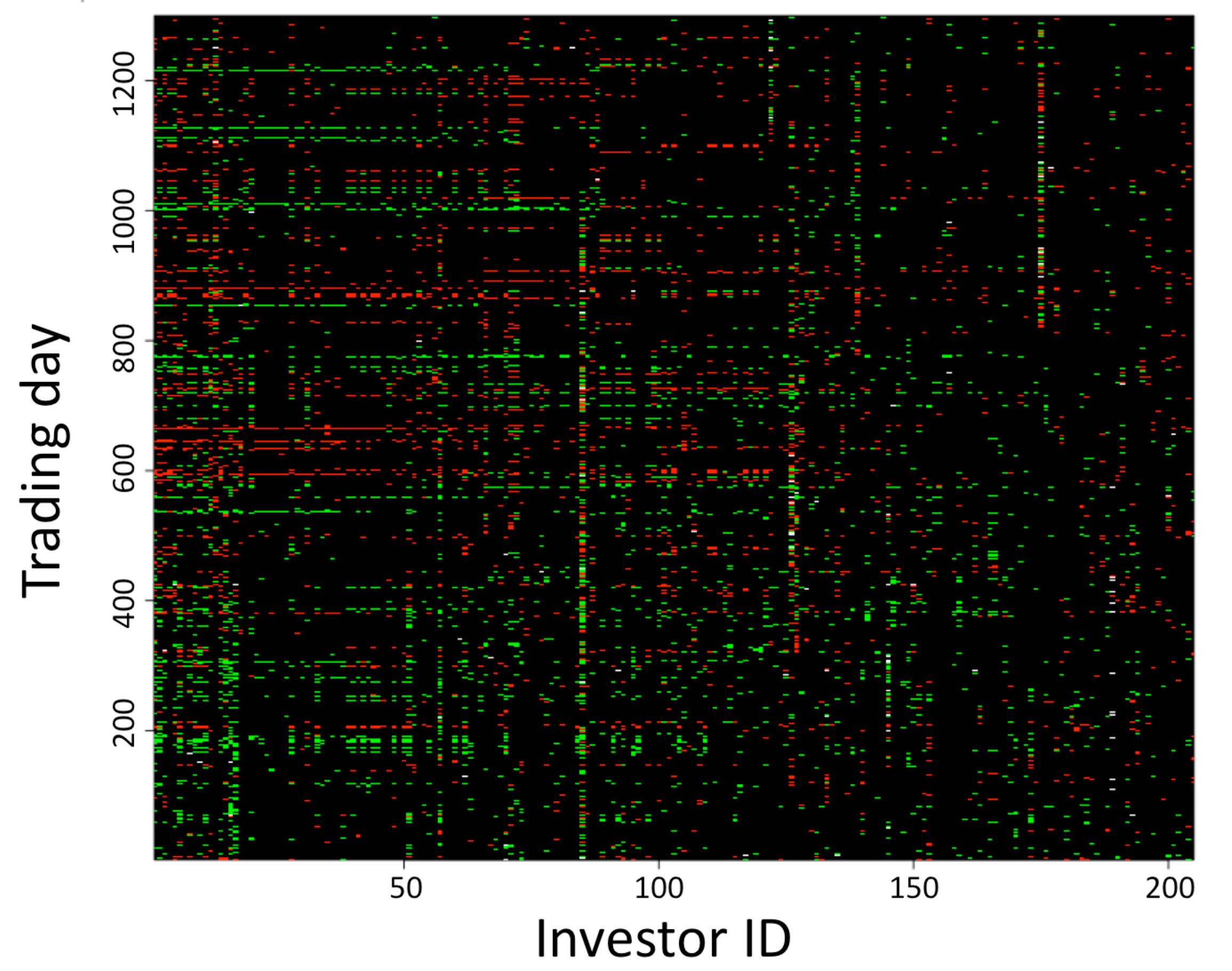}

\includegraphics[scale=0.45]{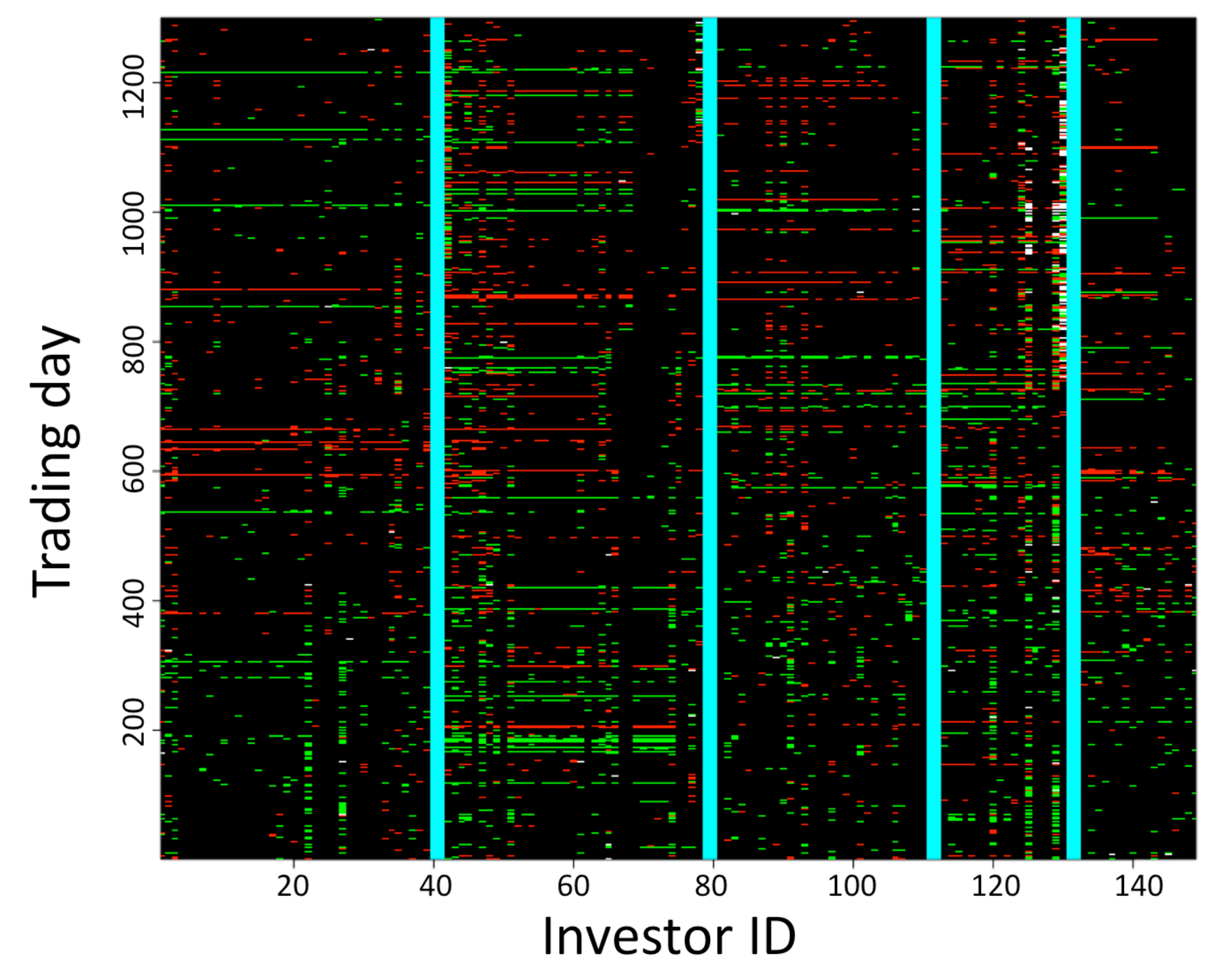}
\caption{Microarray-like representation of the trading activity of investors of the F8 FDR cluster (top panel) and of the B14, B15, B19, B27 and B29 Bonferroni clusters (from left to right bottom panel). Vertical light blue lines separate the clusters.}
\end{center}
\label{ArrayF8}
\end{figure}

Moving from the FDR to the Bonferroni correction, we therefore increase the specificity of the system characterization and decrease its sensitivity. This aspect is summarized in Fig.~\ref{ArrayF8} where we display the trading activities of investors of cluster F8 (top panel) and of clusters  B14, B15, B19, B27 and B29 (bottom panel). Note that 135 elements (out of a total of 141) of the Bonferroni clusters are present in the FDR cluster. The order of the investors in both panels is given according to the rank of the contribution of the single investor to the partitioning of the Infomap algorithm. Highest contribution is provided by investors located at the left of each region. It is worth noting that at the Bonferroni level the specificity of the trading action of each cluster is quite evident and in fact differences among clusters involving trading actions of specific days can be  clearly detected. The FDR cluster provides a less specific characterization but involves a larger number of investors. 

\section{Conclusions}\label{conclusions}

By using a database containing information about the trading actions of individual investors in a real financial market, we have studied how the investors' different actions co-occur in the market. In particular, we have studied the trading of Nokia stock for the period 1998-2003 by associating to each investor for all the days one of the states buy, sell or  buy-sell. Based on the co-occurrence of the trading actions of the investors over time, we have constructed statistically validated networks of investors. This has allowed us also to detect the clusters of investors within the networks and to characterize the observed investors' clusters by the different categories investors belong to, and the type of co-occurence of trading actions, or multi-links, connecting the investors. We have found a very high degree of synchronization in the trading activity of the identified groups. This synchronization can be due to many different causes, such as  the adoption of similar strategies, the recommendation of the same analysts, or a direct interaction and exchange of information among the investors.

Our results demonstrate that despite of the investors' heterogeneity, it is indeed possible and feasible to make empirical observations and characterizations of the investors' actions, to use the concepts and tools of network theory to describe this activity, and to study the clusters of investors  formed in financial markets. The results presented here represent a starting point for further studies focusing on the empirical identification of the investment strategies of the agents \cite{Tumminello2012} and on the modeling of the complex interaction between clusters of agents in a market ecology. We are confident  that  the methods and results presented here will be important in the construction of realistic agent based models of this fascinating complex system. 

\subsection{Acknowledgments}
Authors thank Euroclear Finland (previously Nordic Central Securities Depository Finland) for providing access to the data and MIUR PRIN project 2007TKLTSR  ``Indagine di fatti stilizzati e delle strategie risultanti di agenti e istituzioni osservate in mercati finanziari reali ed artificiali" for financial support.  JP acknowledges the Magnus Ehrnrooth Foundation and the Finnish Academy of Science and Letters - Vilho, Yrj\"o, and Kalle V\"ais\"al\"a Foundation for financial support. MT, FL and RNM acknowledges the NSF project "Financial Markets as an Empirical Laboratory to Study an Evolving Ecology of Human Decision Making" for partial support.


\section*{References}
\begin{thebibliography}{10}

\bibitem{Hommes2006}
Hommes, C H 2006 Heterogeneous Agent Models In Economics and Finance, in {\it Hand-book of Computational Economics, Volume 2: Agent-Based Computational Economics}, edited by L. Tesfatsion and K.L. Judd, (Elsevier Science B.V) 1109-1186 

\bibitem{Samanidou2007}
Samanidou E, Zschischang E, Stauffer D and Lux T 2007 Agent-based models of financial markets {\it Rep. Prog. Phys.} {\bf 70} 409-450

\bibitem{Frankel1990}
Frankel J A and Froot K A 1990 Chartists, Fundamentalists, and Trading in the Foreign Exchange Market {\it Am. Econ. Rev.} {\bf 80} 181-185

\bibitem{Chiarella1992}
Chiarella C 1992 The dynamics of speculative behaviour {\it Ann. Oper. Res.} {\bf 37} 101-123

\bibitem{Kirman1993}
Kirman A 1993 Ants, Rationality, and Recruitment {\it Q. J. Econ.} {\bf 108} 137-156

\bibitem{Lux1995}
Lux T 1995, Herd Behaviour, Bubbles and Crashes {\it Econ. J.} {\bf 105} 881-896

\bibitem{Brock1998}
Brock W A and Hommes C H 1998 Heterogeneous beliefs and routes to chaos in a simple asset pricing model {\it J. Econ. Dyn. Control} {\bf 22} 1235-1274

\bibitem{Lux1999}
Lux T and Marchesi M 1999, Scaling and criticality in a stochastic multi-agent model of financial market {\it Nature} {\bf 397} 498-501 

\bibitem{Chan1988}
Chan K C 1988 On the Contrarian Investment Strategy {\it J. Bus.} {\bf 61} 147-163

\bibitem{Chan1996}
Chan L K C, Jegadeesh N and Lakonishok J 1996 Momentum Strategies {\it J. Financ.} {\bf 51} 1681-1713

\bibitem{Grossman1976} 
Grossman S J and Stiglitz J E 1976 Information and Competitive Price Systems {\it Am. Econ. Rev.}  {\bf 66} 246-253

\bibitem{Roll1984} 
Roll R 1984 A simple implicit measure of the effective bid-ask spread in an efficient market {\it J. Financ.} {\bf 39} 1127-1139

\bibitem{Kyle1985} 
Kyle A S 1985 Continuous auctions and insider trading {\it Econometrica} {\bf 53} 1315-1335

\bibitem{Hasbrouck2007} 
Hasbrouck J 2007 {\it Empirical market microstructure: The institutions, economics and econometrics of securities trading} (Oxford University Press, Oxford, 2007)

\bibitem{Gode1993}
Gode D K and Sunder S 1993 Allocative Efficiency of Markets with Zero-Intelligence Traders: Market as a Partial Substitute for Individual Rationality {\it J. Polit. Econ.} {\bf 101} 119-137

\bibitem{Farmer2005}
Farmer J D, Patelli P and Zovko I I 2005 The predictive power of zero intelligence in financial markets, {\it Proc. Natl. Acad. Sci. USA} {\bf 102} 2254-2259

\bibitem{Nofsinger1999}
Nofsinger J R and Sias R W 1999 Herding and feedback trading by institutional and individual investors {\it J. Financ.} {\bf 54} 2263-2295

\bibitem{Choe1999}
Choe H, Kho B-C and Stulz R M 1999 Do foreign investors destabilize stock market? The Korean experience in 1997
{\it J. Financ. Econ.} {\bf 54} 227-264

\bibitem{Griffin2003}
Griffin J M, Harris J H and Topaloglu S 2003 The dynamics of institutional and individual trading {\it J. Financ.} {\bf 58} 2285-2320

\bibitem{Tumminello2011}
Tumminello M, Miccich\`e S, Lillo F, Piilo J and Mantegna R N 2011 {\it Statistically validated networks in bipartite complex systems} {\it PLoS ONE} {\bf  6}(3) e17994

\bibitem{Tumminello2012} Tumminello M, Lillo F, Piilo J and Mantegna R N 2011 A network description based on co-occurrence of trading decisions of individual investors (manuscript in preparation)

\bibitem{Lillo2008} 
Lillo F, Moro E, Vaglica G and Mantegna R N 2008 Specialization and herding behavior of trading firms in a financial market {\it New Journal of Physics} {\bf 10} 043019

\bibitem{Vaglica2008} 
Vaglica G, Lillo F, Moro E and Mantegna R N 2008 Scaling laws of strategic behavior and size heterogeneity in agent dynamics {\it Phys. Rev. E} {\bf 77} 036110

\bibitem{Moro2009} 
Moro E, Vicente J, Moyano L G, Gerig A, Farmer J D, Vaglica G, Lillo F and Mantegna R N 2009 Market impact and trading profile of hidden orders
in stock markets {\it Phys. Rev. E} {\bf 80} 066102

\bibitem{Toth2010} 
T\'oth B, Lillo F, Farmer J D 2010 Segmentation algorithm for non-stationary compound Poisson processes {\it Eur. Phys. J. B} {\bf 78} 235-243

\bibitem{Carollo2011}  
Carollo A, Vaglica G, Lillo F and Mantegna R N 2011 Trading activity and price impact in parallel markets: SETS vs. off-book market at the London Stock Exchange {\it Preprint at} arXiv:1102.0687

\bibitem{Toth2011} 
T\'oth B, Eisler Z, Lillo F, Bouchaud J-P, Kockelkoren J, Farmer J D 2011 How does the market react to your order flow?  {\it Preprint at}  arXiv:1104.0587

\bibitem {Barber2008}
Barber B M, Lee Y, Liu Y and Odean T 2008  Just How Much Do Individual Investors Lose by Trading? {\it Rev. Fin. Stud.} {\bf 22} 609-632

\bibitem{Kirilenko2010} 
Kirilenko A A, Kyle A S, Samadi M and Tuzun T (2011) The Flash Crash: The Impact of High Frequency Trading on an Electronic Market {\it Available at SSRN}: http://ssrn.com/abstract=1686004

\bibitem{Feiren2011}  
Ren F and Zhou W-X 2011 Analysis of trade packages in Chinese stock market {\it Preprint at}  arXiv:1103.1526

\bibitem{Grinblatt2000}
Grinblatt M and Keloharju M 2000 The investment behavior and performance of various investor types: A study of Finland's unique data set {\it J. Financ. Econ.} {\bf 55} 43-67

\bibitem{Grinblatt2009}
Grinblatt, M and Keloharju M 2009 Sensation Seeking, Overconfidence, and Trading Activity {\it J. Finance} {\bf 64} 549-578

\bibitem{Zipf} 
Zipf G K 1949 {\it Human Behavior and the Principle of Least Effort} (Addison-Wesley, Reading, MA, 1949)

\bibitem{Simon} 
Ijiri Y and Simon H A 1977 {\it Skew Distributions and the Sizes of Business Firms}, (North-Holland, New York, 1977)

\bibitem{Axtell}  
Axtell R L 2001 Zipf distribution of US firm sizes {\it Science} {\bf 293} 1818-1820

\bibitem{Tumminello2011b}
Tumminello M {\it et al.} 2011 Happy aged people are all alike, while every unhappy aged person is unhappy in its own way, {\it to appear in PLoS ONE}

\bibitem{Miller1981}
Miller R G 1981 {\it Simultaneous Statistical Inference} (2nd ed. New York: Springer-Verlag, 1981)

\bibitem{Benjamini1995}
Benjamini Y and Hochberg Y 1995 Controlling the False Discovery Rate: a practical and powerful approach to multiple testing {\it J. R. Statist. Soc. B} {\bf 57} 289-300

\bibitem{Rosvall2008}
Rosvall M, Bergstrom CT 2008 Maps of random walks on complex networks reveal community structure {\it Proc. Natl. Acad. Sci. USA} {\bf 105} 1118-1123

\bibitem{Lancichinetti2009}
Lancichinetti A and Fortunato S 2009 Community detection algorithms: A comparative analysis.
{\it Phys Rev E} {\bf 80} 056117

\bibitem{Jstat} 
Tumminello M, Miccich\`e S, Lillo F, Varho J, Piilo J, Mantegna R N 2011 Community characterization of heterogeneous complex systems {\it J. Stat. Mech-Theory Exp.} P01019

\end{thebibliography}
\end{document}